\documentclass[a4paper,11pt,reqno]{amsart}
\usepackage[body={15.5cm,20.3cm}]{geometry}
\usepackage{amssymb}
\usepackage{hyperref}
\usepackage{enumerate}

\newcommand*{\mailto}[1]{\href{mailto:#1}{\nolinkurl{#1}}}

\theoremstyle{plain}
\newtheorem{thm}{Theorem}[section]

\theoremstyle{remark}
\newtheorem*{rem*}{Remark}
\newtheorem*{rems*}{Remarks}
\theoremstyle{definition}

\newtheorem*{assus*}{Assumptions}
\newtheorem*{Def*}{Definition}
\newtheorem*{not*}{Notation}

\providecommand{\C}{\mathcal}
\providecommand{\D}{\mathbb}
\newcommand{\bR}{\D{R}}

\newcommand{\bE}{\D{E}}

\DeclareMathOperator{\prob}{\D{P}}

\DeclareMathOperator{\tr}{tr}

\newcommand{\Hmm}[1]{\leavevmode{\marginpar{\tiny%
$\hbox to 0mm{\hspace*{-0.5mm}$\leftarrow$\hss}%
\vcenter{\vrule depth 0.1mm height 0.1mm width \the\marginparwidth}%
\hbox to 0mm{\hss$\rightarrow$\hspace*{-0.5mm}}$\\\relax\raggedright #1}}}

\begin{document}

\title[Erratum]{Erratum to: From Uncertainty
Principles to Wegner Estimates}

\author[P. Stollmann]{Peter Stollmann$^{\dagger}$}
\address{$^{\ddagger}$Fakult\"at f\"ur
   Mathematik, Technische Universit\"at, 09107 Chemnitz, Germany}
\email{\mailto{peter.stollmann@mathematik.tu-chemnitz.de}}
\date{April 1, 2011}
\maketitle
\section*{The error and what remains true}
In March 2011 Abel Klein pointed out that the proof of Theorem 4.1 in my paper \cite{sto} suffers from a fatal mistake. In fact, what Abel rightly mentioned is that the second inequality sign in (4.5) is not supported by any argument and is probably wrong. Here and in the following we refer to the numbering in \cite{sto} without further mentioning. The following still holds:

\begin{thm}
\label{wegner}
Let $H$ be as in the setup, $W:= \sum_{\alpha\in\C{I}} U_\alpha$ and $I$ be some interval in $\bR$. Assume that there is $\kappa>0$ such that $\prob$-a.s.
 \begin{equation}      \label{eq:uncertainty}
P_I(H)WP_I(H)\geq\kappa P_I(H). \tag{$\star$}
\end{equation}
Then, there is a constant $C$ such that
\begin{equation}\label{eq:wegner}
 \bE\{\tr[P_I(H)]\}\le C |\Lambda|s(\prob,| I|) .
\end{equation}
\end{thm}
The proof had been given by Combes, Hislop and Klopp in \cite{CHK-07}, proof of Thm 1.1 and we stress the fact that in this part of their work no periodicity assumptions are used. The main difference to the claim in \cite{sto} is that we loose the very explicit control of the constant; apart from that all the results remain valid.

\subsection*{Acknowledgement}
Sincere thanks go to Abel Klein, who pointed out the error in the original version of the proof and bewared me of further wrong versions :)

\end{document}